\documentclass[baaa]{baaa}

\usepackage[pdftex]{hyperref}
\usepackage{subfigure}
\usepackage{natbib}
\usepackage{helvet,soul}
\usepackage[font=small]{caption}

\contriblanguage{1}

\contribtype{1}

\thematicarea{2}

\received{\ldots}
\accepted{\ldots}

\title{Hurst exponent and planetary rings}

\titlerunning{Hurst exponent and planetary rings}

\author{H. Salomone\inst{1} \& N.E.  Grandi\inst{2,3}}
\authorrunning{Salomone et al.}

\contact{hsalomon@campus.ungs.edu.ar}

\institute{
Instituto de Industria, UNGS, Argentina \and
Instituto de Física de La Plata, CONICET--UNLP, Argentina \and
Departamento de Física, UNLP, Argentina
}

\resumen{
Verificamos 
en los anillos de Saturno
si el exponente Hurst es un observable “robusto”, en el sentido de que devuelve valores consistentes para diferentes observaciones del mismo objeto.
}

\abstract{
We test 
on Saturn's rings
whether the Hurst exponent is a “robust” observable, in the sense that it returns consistent values for different observations of the same object. 
}


\begin{document}

\maketitle

\section{Introduction}
\label{S_intro}
All four giant planets in our Solar System posses some kind of ring structure around them, the largest and most visible being those of Saturn. These rings are formed by particles whose sizes span from micrometers to meters. While predominantly composed of water ice, the ring particles also contain minor traces of rocky material (for a review, see \cite{colwell2009structure}).

Planetary rings exhibit fractality, consisting in detailed features at multiple scales. This motivated the original suggestion of \cite{mandelbrot1982fractal} that they might be well described by a Cantor-like set. A theoretical motivation for this hypothesis was proposed in \cite{avron1981almost}, when the pictures taken by Voyager missions started to unveil the complexity of Saturn's ring system. Later, the Cassini mission provided a large number of high resolution pictures contributing to a better understanding of the rings. In this context, \cite{li2015edges} proposed to assign a fractal dimension to the edges of the rings (see also  \cite{malyarenko2017fractal}). In these works, the identification of the edges of the rings involved the intensification of the contrast in order to obtain a black and white image, to which the box-counting technique can be applied.

It is the aim of these article to broad the aforementioned fractal analysis, in order to take into account the brightness of Saturn's rings as captured the images of Cassini. Our proposal is to calculate the Hurst exponent of the series corresponding to the grayscale value of the ring pictures as a function of the radius.
Originally proposed by the British hydrologist Harold Edwin Hurst, such exponent is a statistical observable that measures the long-range dependencies and self-similarity on a series. 
It is the aim of this work to test whether the Hurst exponent is a ``robust'' observable, in the sense that it returns consistent values for different observations of the same object. 


\section{Hurst exponent}
The Hurst exponent can be measured on a one-dimensional series of observations $X_i$ with a total of $N$ components
\begin{equation}
X_1,X_2,\dots,X_N
\label{eq:X}
\end{equation}
To define it, the first step is to take any subseries of length $n<N$, and write it in terms of the standarized values
\begin{equation}
Y_i=\frac{X_{i}-m(n)}{S(n)} \qquad \qquad \text{  for } i=1,2, \dots ,n \,,
\end{equation}
which is defined in terms of the mean value of the subseries $m=\sum_{i=1}^{n} X_i/n$ and its standard deviation $S^2(n)= \sum_{i=1}^{n}\left ( X_{i} - m \right )^{2}/n\,$. The cumulative deviate of the subseries at the point $j$ is then defined as the partial sum
\begin{equation}
Z_j = \sum_{i=1}^{j} Y_{i} \qquad\qquad\qquad \text{   for }  j=1,2, \dots ,n \,. 
\end{equation}
This represents the ``area under the curve'' from the first element of the series up to the $j$-th value. Next, the \emph{rescaled range} is defined as
\begin{equation}
R(n) =\operatorname{max}\left (Z_1,   \dots, Z_n  \right )-
  \operatorname{min}\left (Z_1, \dots, Z_n  \right ).
\end{equation}
Notice that a small value for $R(n)$ implies a rapidly oscillating series, while a large one corresponds to a more smooth behavior. This is naturally a growing function of the length $n$ of the subseries. In the next step, we average over all the possible subseries of $n$ subsequent values to obtain the averaged rescaled range as $\langle R(n)\rangle$, and then define the Hurst exponent according to
\begin{equation}
\langle R(n)\rangle \propto n^H
\label{eq:H}
\end{equation}
This growth is slower than linear, implying that $H$ is in the range $0<H<1$. A large $H$ implies a more regular series, while a small $H$ signals an erratic one. In this sense, $H$ is a good characterization of the fractality of the series.

\begin{figure*}[!t]
  \centering
  \includegraphics[width=0.42\textwidth]{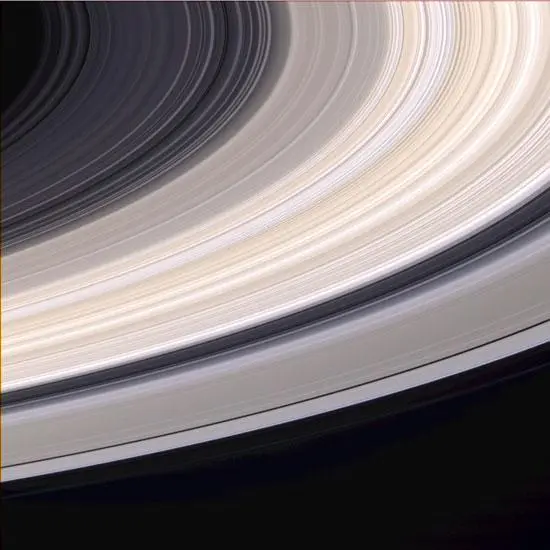}  \quad
  \includegraphics[width=0.42\textwidth]{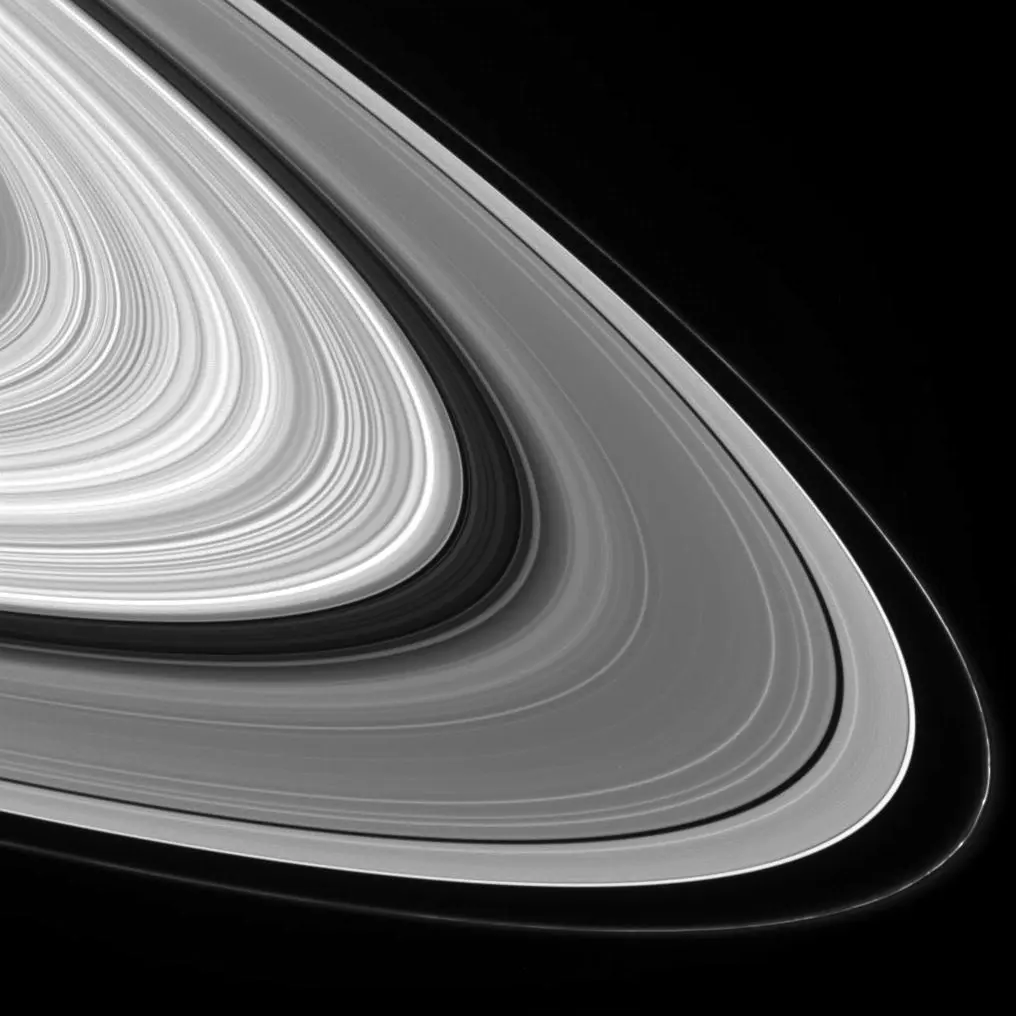}
  \\ ~\\ 
  \includegraphics[width=0.42\textwidth]{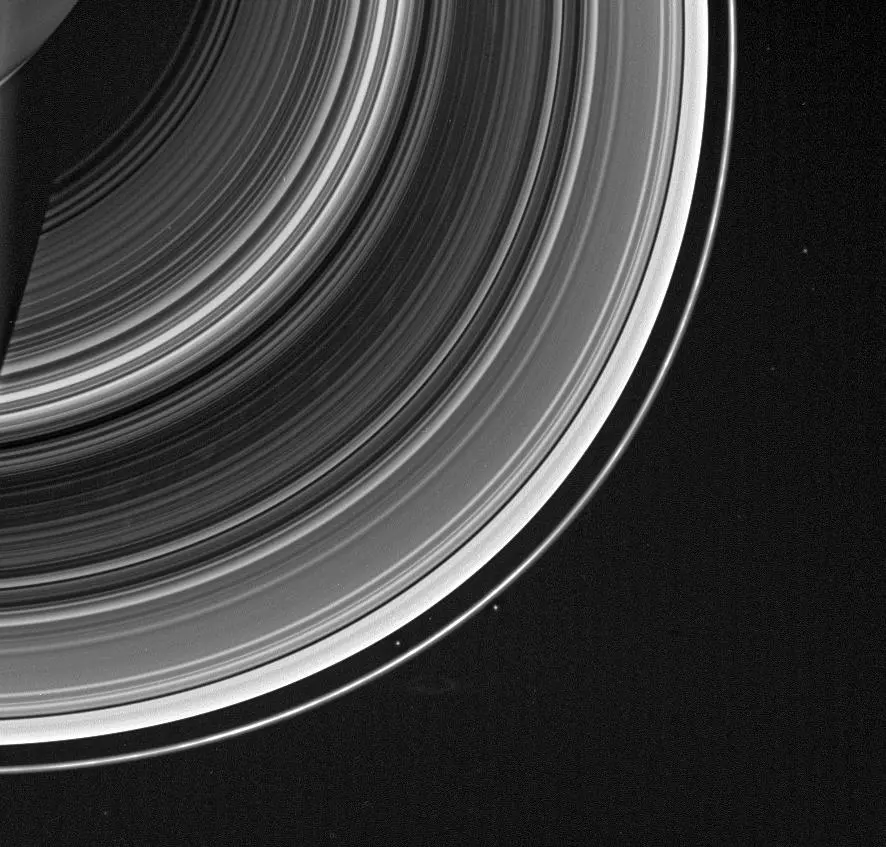}  \quad
  \includegraphics[width=0.42\textwidth]{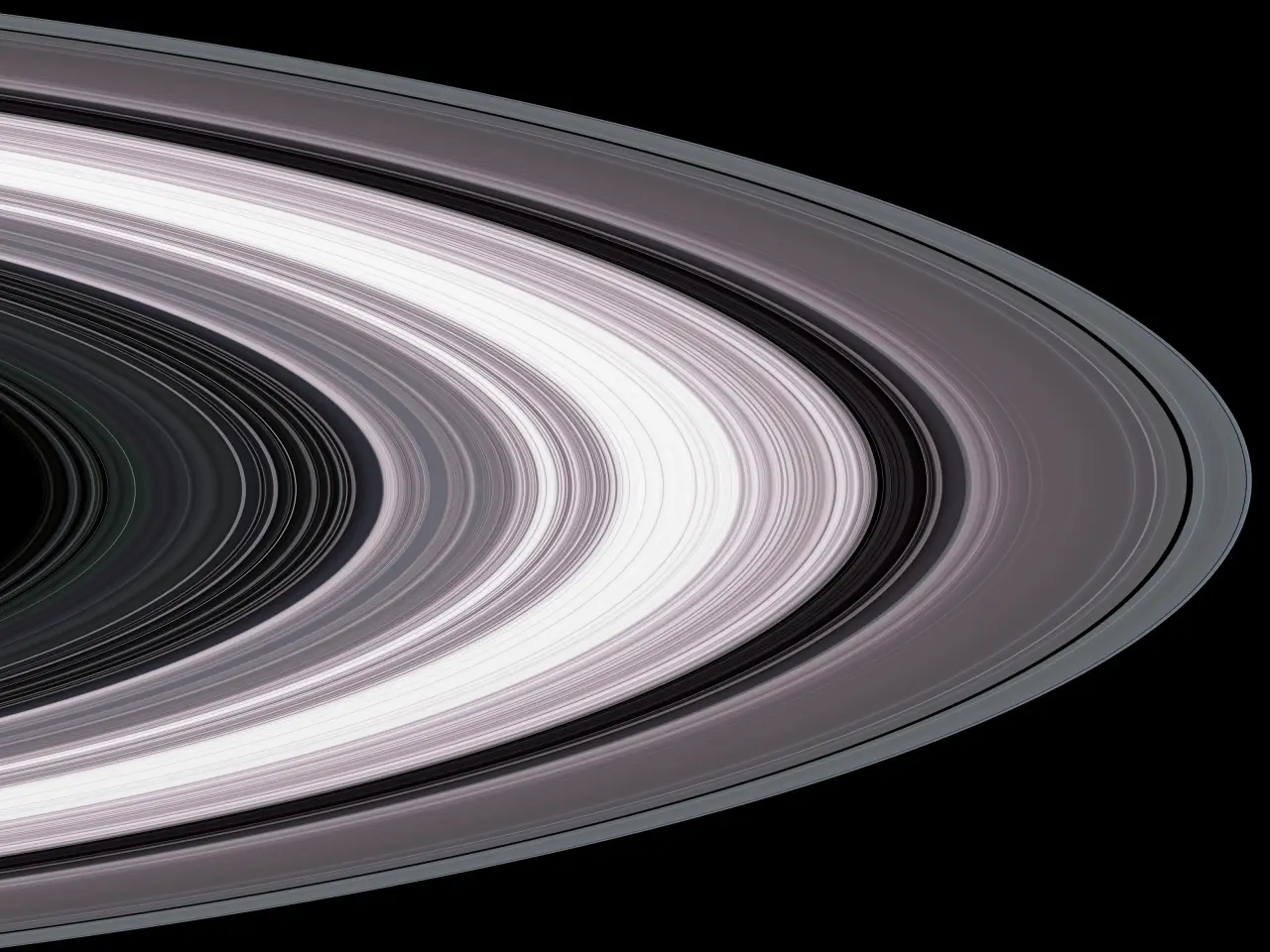}
  \caption{Example pictures taken from the set used to calculate the Hurst exponent. Source: NASA Cassini-Huygens mission website, the images are identified as PIA05421 (top left), PIA08275 (top right), PIA08859 (bottom left) and PIA07872 (bottom right).}
  \label{fig:pictures}
\end{figure*}
 
\section{Application to Saturn's rings}
To apply the Hurst exponent to Saturn's rings, we select a subset of 15 photos from the Cassini-Huygens mission grayscale images, taken from NASA website\footnote{The images are those identified as PIA05396, PIA07631, PIA07872, PIA08253, PIA08275, PIA08859, PIA08872, PIA08937, PIA09763, PIA11466, PIA11657, PIA20499, PIA22418, from the site \url{https://www.jpl.nasa.gov}.}, 
some examples of which are shown in Fig.~\ref{fig:pictures}. 
 
\begin{figure*}[!t]
  \centering
  \includegraphics[width=0.67\textwidth]{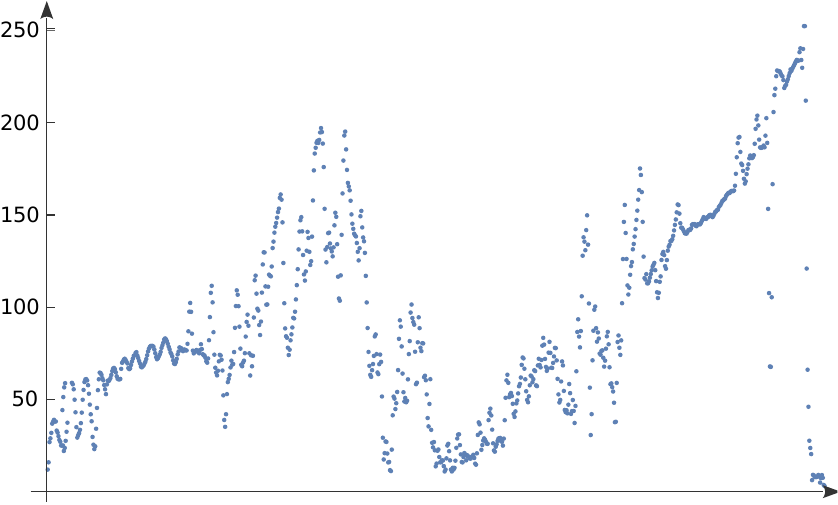}
  \caption{
Example of a grayscale series representing the brightness of Saturn's rings as a function of the radius. This correspond to the NASA Cassini-Huygens mission identified as PIA08859, bottom left of Fig.~\ref{fig:pictures}.}
  \label{fig:serie}
\end{figure*}

We define the series $X$ in Eq.  \ref{eq:X} as the grayscale value of the image as a function of the radius, measured from the interior edge of the ring system, at a fixed orbital angle, up to its exterior edge. 
As can be seen in Fig.~\ref{fig:serie}, this function presents an oscillatory behavior in a wide range of scales. This makes it a natural arena to explore its characterization via the Hurst exponent.


We use a {\tt MATLAB} code to systematically process the images\footnote{The code is available at \url{https://git.nixnet.services/nicolasg/Hurst-exponent.git}}. The algorithm performs the following steps:
1) extracts grayscale values from  800 radial slices of the rings for each image, 2) removes non-informative black regions (inner and outer edges) to isolate the relevant data, 3) normalizes the radial coordinate to compress the profile into a uniform radial range, and finally 4) computes the Hurst exponent $H$ using Eq. \ref{eq:H}.

We obtain a very consistent value for $H$ across different angles in the same picture and across the 15 different pictures. An important observation is that the  Hurst exponent is not sensitive to how aligned is the slice with the radius of the rings, as long as it goes from the innermost edge to the outermost one. 
We get a mean value and a variance for the Hurst exponent of $H=0.93\pm 0.04$.

\section{Discussion}

The observed dispersion in our analysis constitutes less than $5$\% of the mean value, indicating that the Hurst exponent \( H \) serves as a robust metric for characterizing the structural properties of Saturn’s rings. However, it is important to emphasize that this observable was initially developed to quantify granularity, rendering the results presented here inherently sensitive to the spatial resolution of the input images. Consequently, the reliability of \( H \) as a robust observable can only be rigorously assessed once a well-defined resolution threshold is established.  

A primary limitation of this study concerns the image data used. Sourced from platforms designed for visual representation, these images cannot guarantee unaltered fidelity. Furthermore, image brightness is not a reliable indicator of optical depth, as it is influenced by variable illumination angles (dependent on the Sun–rings–Cassini probe geometry) and differences in albedo across ring regions. A more sensible alternative involves optical depth profiles derived from stellar occultation measurements\footnote{Available via \url{https://pds-rings.seti.org/voyager/pps/profiles.html}}.

For occultation data at $1km$ resolution, we obtain a Hurst exponent of \( H = 0.87^{+0.02}_{-0.07} \). Direct comparison to our results is initially precluded by resolution disparities: our image-derived resolution of $~70km$ (estimated by dividing the $~70,000km$ ring system width by $1024$ pixels) significantly exceeds the $1km$ occultation data resolution. To address this discrepancy, we convolve the high-resolution occultation data with a Gaussian kernel of $70km$ width to match our effective resolution. Strikingly, this adjusted dataset yields \( H = 0.93^{+0.002}_{-0.004} \), in very good agreement with our original result of \( H = 0.93 \pm 0.04 \).

A more ambitious program would entail the exploration of the Hurst exponents for the rings of the other giant planets of our Solar System. A preliminary calculation of the Hurst exponent for a single picture of the Uranus's rings taken by Voyager 2\footnote{Images are taken from \url{https://jpl.nasa.gov/images/pia00142-uranus-ring-system}}  gives $H=0.88$. However, since in this case the resolution is not quite good, this result could correspond to artifacts of the image rather than a property of the rings themselves. The result obtained from the occultation data with $1km$ resolution is $H=0.91\pm0.04$. By adjusting the resolution to that of the picture following the procedure explained above, we get $H=0.90\pm0.04$.

The expected value for an uncorrelated series is $H=0.5$. A series with larger values $H>0.5$ tends to have positive self-correlation, high values tend to be followed by high values. On the other hand, a series with smaller values $H<0.5$ corresponds to events with alternating correlation, high values tend to be followed by low values. Our results show that Saturn's rings have positive self-correlation in the radial direction: dense regions tend to be followed by dense regions.

\begin{acknowledgement}
We thank Guillermo Silva, Rodolfo Echarri and Pablo Pisani for helpful comments and encouragement. This work was partially supported by  CONICET grant PIP-2023-11220220100262CO, UNLP grant 2022-11/X931, and UNGS grant 2022-30/4138.
The Hurst analysis presented here was part of a MSs thesis in the {``Maestría en Física Contemporánea''} of La Plata University. 
\end{acknowledgement}


\bibliographystyle{baaa}
\small
\bibliography{bibliografia}
 
\end{document}